\documentclass[aps, prb, amssymb,amsmath,reprint]{revtex4-1}
\usepackage{graphicx}
\usepackage{textcomp}
\usepackage[update]{epstopdf}

\begin{document}
\title{Metallic State of Low Mobility Silicon at High Carrier density induced by an Ionic Liquid}

\author{JJ \surname{Nelson}}
\email{nelson@physics.umn.edu}

\author{A. M. \surname{Goldman}}
\email{goldman@physics.umn.edu}

\affiliation{School of Physics and Astronomy, University of Minnesota, 116 Church Street SE, Minneapolis, Minnesota 55455, USA}

\date{\today}

\begin{abstract}
High mobility and dilute two-dimensional electron systems exhibit metallic behavior down to the lowest experimental temperatures. In studies of ionic liquid gated insulating silicon, we have observed transitions to a metallic state in low mobility samples at much higher areal carrier densities than found for samples of high mobility. We have also observed a mobility peak in metallic samples as the carrier density was increased beyond $10^{13} \text{cm}^{-2}$.
\end{abstract}



\maketitle

It was initially believed that a metallic regime would not be found
in two dimensional (2D) systems. As a function of carrier density
there would only be a crossover from weak to strong localization.\citep{Abrahams(1979)}
This scenario was confirmed experimentally in highly disordered silicon
systems almost 30 years ago, with no metallic state being found. As
sample mobilities increased, metallic regimes were observed. In the
case of n-type Si MOSFETs with mobility $\mu\sim10^{4}\text{ cm}^{2}\text{/Vs}$
metallic behavior was found at $n_{c}\sim10^{11}\text{ cm}^{-2}$.\citep{Kravchenko(1994)}
In \textit{n} and \textit{p} -type GaAs/AlGaAs quantum wells with
$\mu\sim10^{6}\text{ cm}^{2}\text{/Vs}$, a transition was found at
$n_{c}\sim10^{10}\text{ cm}^{-2}$\citep{Hanein(1998),Dultz(1998)}.
There is currently a question as to whether these metal-insulator
transitions (MITs) are crossovers or quantum phase transitions. \citep{DasSarma(2014)Sept}
A thorough review of the experiments can be found in Ref. \onlinecite{Spivak(2010)}. In addition to exhibiting
2D MITs, \textlangle100\textrangle{} Si has been shown to be superconducting
with heavy B doping and \textlangle111\textrangle{} Si wafers have
been shown to be superconducting with monolayers of Pb or In on the
crystal surface.\citep{Bustarret2006,Zhang2010} The search for superconductivity
in Si using ionic liquid gating was the original motivation for this
work.

With the use of ionic liquid gating, we have created 2D hole gases
(2DHGs) on low mobility Si wafers. In this Letter we report both a
transition to a metallic state with a critical sheet carrier density
higher than previously reported and a peak in the mobility as a function
of carrier density. A high carrier density metallic state, while unexpected,
was predicted to exist by Das Sarma and Hwang.\citep{DasSarma(2014)2D}
The goal of their theory was to determine a scaling relationship between
the critical carrier density to observe metallic behavior and the
peak mobility to explain the different critical carrier densities
observed in MIT experiments. This was done by looking at the 2D conductivity
using the Drude model in which it is determined by the product $k_{f}l$
where $k_{f}$ is the Fermi wavevector and $l$ the mean free path.
Boltzmann transport theory was used to determine the dependence of
$l$ on carrier density $n$ and the Ioffe-Regel criteria ($k_{f}l=1$)
was used to determine the carrier density where a crossover to metallic
behavior would be observed. In their theory the transition to a metallic
state is not a phase transition but is a crossover. 

Ionic liquids (ILs) such as N,N-diethyl-N-(2-methoxyethyl)-N-methylammonium
bis(trifluoromethylsulphonyl-imide) (DEME-TFSI) have been used to
study MITs and superconductor-insulator transitions by electrostatic
charging because using them one can reach higher carrier concentrations
than achievable with conventional capacitor configurations employing
solid dielectrics.\citep{APLMisra(2007),Shimotani(2007)} Ionic liquids
are in effect room temperature molten salts. They consist of mobile
charged ions that will move to the electrode surfaces with an applied
electric field. Electric double layers form at the positive and negative
electrodes with the bulk liquid remaining in a neutrally charged state.
These ion/electrode interfaces can be viewed as individual capacitors
with nanometer level separation. High carrier concentrations are achievable
without the limitations of leakage due to electron tunneling or dielectric
breakdown characteristic of devices using thin dielectrics. Gating
of oxides with an ionic liquid under certain conditions is also known
to induce oxygen vacancies.\citep{Jeong22032013} The current understanding
is that the process of inducing charge carriers with ionic liquid
based transistors may involve both electrostatic and chemical processes
with the balance between the two determined by voltage and temperature.
Gating at low temperatures can inhibit electrochemical reactions,
but gating must be performed at high enough temperatures that the
ions remain mobile. 

\begin{figure}
\includegraphics[width=8.5cm]{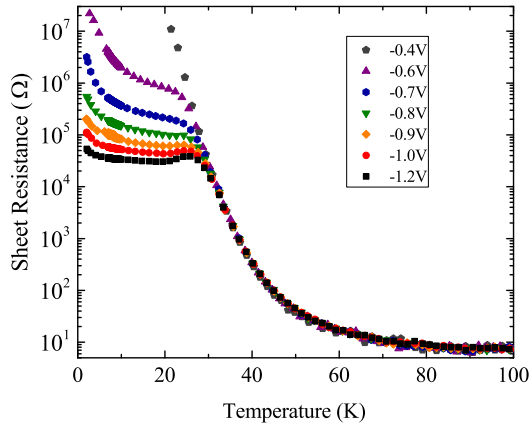}\protect\caption{\label{fig:1}(color online) Sheet resistance vs temperature for ionic
liquid gated silicon. The data were taken as a function of time during
the cool down. The figure demonstrates that gating does not appear
to change the bulk behavior but results in a low temperature surface
conducting state. The conducting state is sensitive to the surface
passivation, roughness, and crystal orientation, which may vary from
sample to sample.}
\end{figure}

Samples were prepared from Si wafers grown by the Czochralski process.
The numbers of oxygen inclusions are larger and the mobilities lower,
with wafers grown this way, as compared with float-zone grown wafers.\citep{Borghesi(1995)}
The wafers were B doped and exhibited resistivities of $1-5\text{\ensuremath{\Omega\text{cm}}}$.
Silicon MOSFETs used to observe the MIT employed \textit{pn} junctions
to isolate the conductive channel from the bulk wafer. In the present
work we induced additional holes onto the surface of a \textit{p}-type
Si wafer. Silicon exhibits an activated resistance as a function of
temperature due thermal excitation of carriers above an energy gap.
Conduction in a surface channel should become observable at temperatures
below those at which the bulk carriers freeze out. The channel sheet
resistance was determined using the van der Pauw method\citep{SupDoc}. 

Trapped charges in the oxide coating can spontaneously form an inversion
layer on a \textit{p}-type Si wafer, and if such a layer were present,
the 2K resistance could be as low as a $\text{M\ensuremath{\Omega}}$.
While rare, this was observed in several samples, but the behavior
was suppressed with the accumulation of holes on the surface. At 2K
the resistance of the Si samples would be above the input impedance
of the electronics unless a conductive surface layer were formed.
When increasing the magnitude of the gate voltage, $V_{g}$, we observe
no change at high temperatures, which we attribute to the dominance
of bulk conduction, but we do observe a decrease in the low temperature
resistance from $>10\text{M\ensuremath{\Omega}}$ to $\text{k\ensuremath{\Omega}}$,
as shown in Fig. \ref{fig:1}. We have only been able to produce such
a state by gating of \textlangle100\textrangle{} Si wafers. We could
not do this using \textlangle111\textrangle{} Si wafers with a similar
B doping level. The results are reminiscent of those of Ando \textit{et
al}. in which $\text{Na}^{+}$ ions trapped within the $\text{SiO}_{2}$
layer were reported to form a 2D surface state in p-type Si.\citep{Ando(1982)}

\begin{figure}
\includegraphics[width=8.5cm]{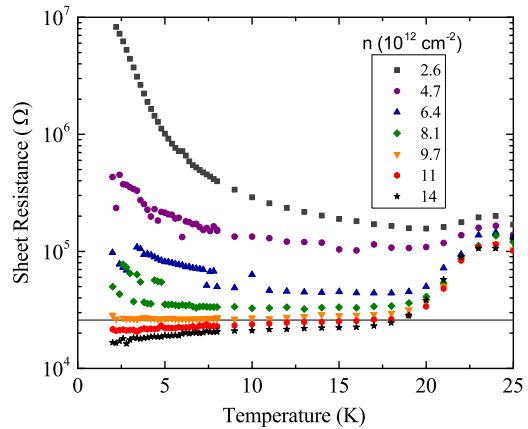}\protect\caption{\label{fig:2}(color online) Sheet resistance vs temperature of an
ionic liquid gated silicon sample that shows metallic behavior. The
inset is labeled with carrier densities in units of $10^{12}cm^{-2}$
and below 15K a MIT can be seen in the conducting surface layer in
this sample. The horizontal line is drawn at a resistance of $h/e^{2}$.}
\end{figure}

Carrier concentrations of samples were obtained in two different ways,
from the Hall effect at low temperatures and by integrating the current
flowing to the gate during charging. The carrier concentrations in
Fig. \ref{fig:2} were determined using the latter procedure. The
carrier density produced by the integration method is also a measure
of the number of excess anions that moved to the $\text{SiO}_{2}$
surface. In this approach there is a nonzero leakage current that
needs to be accounted for, and this was done by subtracting off the
value of the current in the long time limit. The subtracted term was
at the level of, or below a nA, and lead to a small change in the
integrated area. Integration resulted in carrier densities within
a factor of order unity to those obtained from the Hall number. This
method can fail at higher voltages where the leakage current is large
and does not asympototicly decay. When oxidation of the electrode
occurs, the integrated carrier density can be a factor of 10 higher
than that of the Hall carrier density.\citep{Petach(2014)} Our experiments
were designed to minimize oxidation processes.\citep{SupDoc} 

The Hall effect was measured at $2\text{K}$ for some samples and
from the Hall number, $n\sim10^{12}\text{cm}^{-2}$ and $\mu\sim20\text{cm}^{2}\text{/Vs}$
when the sheet resistance, $R_{\square}(2\text{K})\sim10^{4}\Omega$.
Many samples exhibited measurable conduction at low temperatures but
only a few showed metallic behavior. Samples that exhibited metallic
behavior remained metallic down to 450mK. This was the lowest accessible
temperature. The resistance vs. temperature of the insulating state
at low hole concentrations was Arrhenius activated, and it transitioned
to a regime in which $\sigma\sim\ln(T)$ at higher hole concentrations
as the metallic regime was approached.

An example of a sample that showed metallic behavior is plotted in
Fig. \ref{fig:2}. The data were collected by warming the sample and
the lowest measured temperature was 2K. High mobility silicon MOSFETs
exhibited a transition to a metallic state at a resistance of $3h/e^{2}$.
The present samples exhibited a transition at $h/e^{2}$, a value
also reported by Zataritskaya and Zavaritskaya for p-type and n-type
Si inversion layers.\citep{Zataritskaya(1987)}

Ionic liquid gating of Si was determined to be reversible without
hysteresis across the MIT.\citep{SupDoc} This is in contrast with
the MIT observed in $\text{VO}_{2}$ where the samples remained metallic
when the negative gate voltage was removed.\citep{Jeong22032013}
Oxidation of the $\text{VO}_{2}$ sample was believed to be the source
of the hysteresis in the experiment. In addition, a chemical reaction,
if it occurred reversibly, would happen at the interface between the
ionic liquid and the silicon oxide, whereas the conductive channel
is a distance away, spatially separated by the thickness of the oxide.

Das Sarma and Hwang predicted that the carrier density, $n_{c1}$,
at the initial transition to metallic behavior would scale with the
inverse of the peak mobility.\citep{DasSarma(2014)2D} For the values
of the mobilities of the wafers of the present work, they would predict
that $n_{c}\sim10^{12}\text{cm}^{-2}$; however, it was believed that
such systems would be too disordered for a metallic state to exist.\citep{DasSarma(2014)2D}

The observation of a transition to a metallic state in low mobility
devices was unexpected, as previous findings all involved high mobility
samples. An open question remains as to whether the high carrier density,
low mobility metallic state is similar to the low carrier density
high mobility state studied by others and whether the transition from
the insulating to metallic state is a quantum phase transition or
a crossover. At present we have no answer to these questions. Long
range interactions dominate at the carrier densities of previous experiments
whereas the low mobility transition occurs at a carrier concentration
where short range interactions should be an important factor. 

A measure of the importance of long-range interactions is the dimensionless
parameter $r_{s}^{-1}=(\pi n_{s})^{\frac{1}{2}}a_{B}$ where $a_{B}$
is the Bohr radius. This parameter is the ratio of the Coulomb energy
to the Fermi energy. For dilute 2D systems, $r_{s}$ is typically
high at the critical carrier density for the MIT. Using the bulk parameters
for Si we calculate $r_{s}\sim5$ for the heavy hole band. This suggests
that the metallic regime of these ionic liquid gated systems could
be different from those of MOSFETs. However a calculation using Si
band parameters might not be the whole story. In MOSFETs the metallic
gates must be far enough away from the channel to suppress electron
tunneling from the gate. With ionic liquid gating, the ions sit on
the oxide surface $1-2\text{ nm}$ from the 2D electron gas. We cannot
rule out the possibility that the ions produce a strong spatially
varying electric field that the carriers in the conductive channel
experience. The interaction of accumulated holes with this potential
could lead to an increase in the hole effective mass, a decrease in
$a_{B}$, and thus an increase in $r_{s}$ from the values calculated
from silicon band parameters.

\begin{figure}
\includegraphics[width=8.5cm]{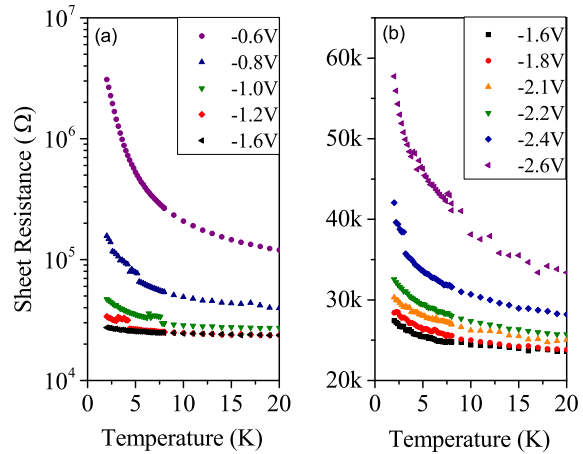}\protect\caption{\label{fig:3}(color online) (a) Semilog plot of sheet resistance
vs. temperature. The sample becomes more conducting as holes are added.
(b) Linear plot of resistance vs. temperature of the same sample.
The sample exhibits increased insulating behavior as holes are increased
further. }
\end{figure}

Previous experiments on low mobility samples at very high carrier
density observed that electron mobility decreased with increasing
carrier density and there was no metallic regime.\citep{Ando(1982)}
The decrease in carrier mobility is typically associated with either
surface surface roughness or scattering from trapped charges in the
oxide. Similar observations have been made with electrolyte gated
rubrene samples where the decrease in mobility was thought to be due
to electrostatic disorder produced by the ions.\citep{Wei(2014)}
A peak in the mobility as a function of carrier density was also observed
in the $\text{Si}$ surface layer presented here. A typical result
is plotted in Fig. \ref{fig:3}, where the sample became less insulating
with increasing carrier density at low carrier densities and more
insulating at higher carrier densities, implying the existence of
a peak in the mobility at a carrier density of the order of $10^{13}$$cm^{-2}$
as measured by the Hall effect. Wei \textit{et al}. have reported
a similar peak in gated rubrene at similar carrier densities.\citep{Wei(2014)}
In addition, the data such as that of Fig \ref{fig:3} were repeatable
when the sample was removed from the system and attached to a low
temperature insert for a subsequent run. Other samples were similarly
reproducible and this suggests that degradation of the sample is not
an explanation of the high carrier density behavior. In contrast with
Fig. \ref{fig:2}, metallic behavior was not observed in the sample
plotted in Fig. \ref{fig:3}. Metallic samples also show a peak in
the mobility as a function of carrier density. 

\begin{figure}
\includegraphics[width=8.5cm]{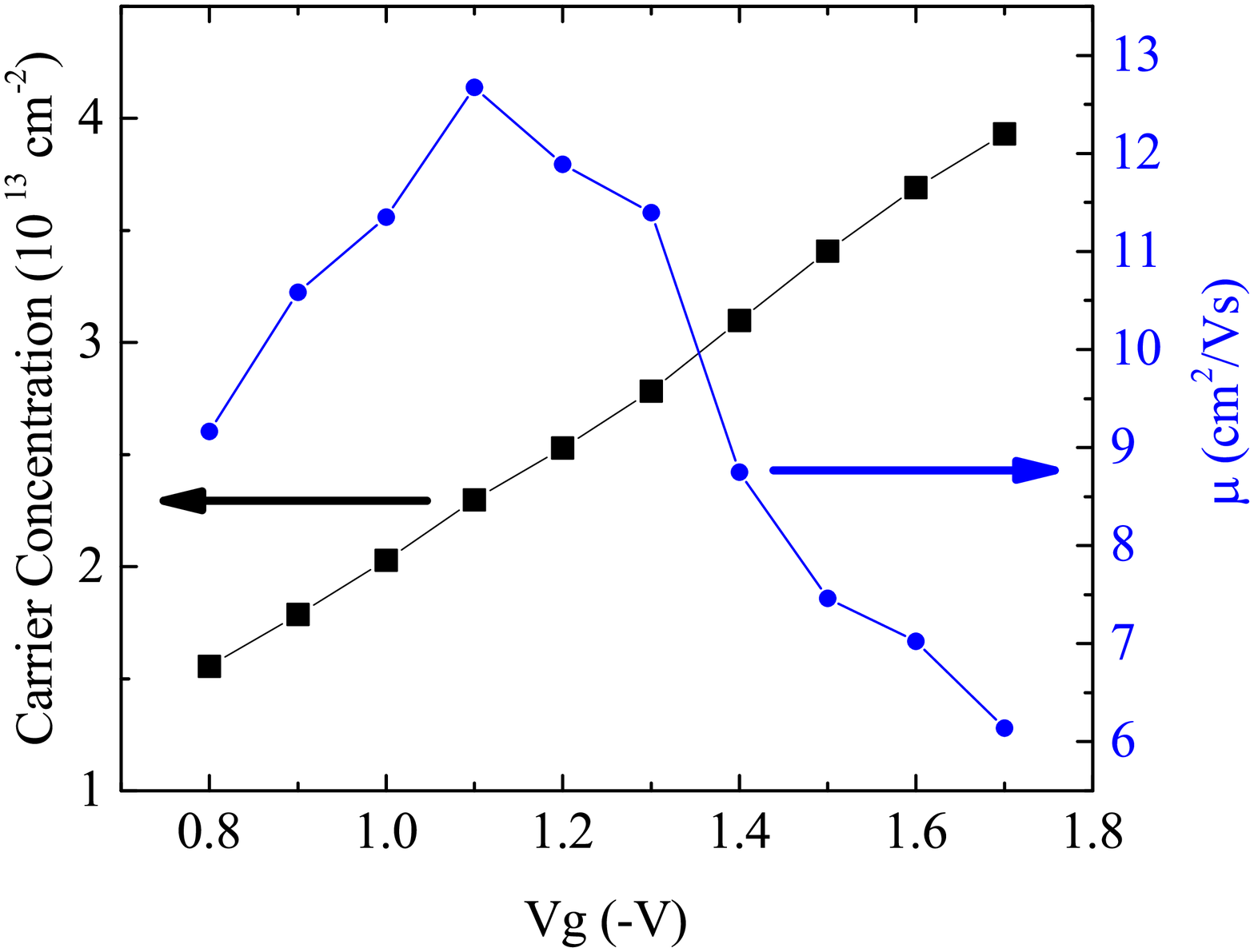}\protect\caption{\label{fig:4}(color online) Integrated carrier concentration and
mobility as a function of gate voltage. The mobility was calculated
from the conductivity at $2\text{K}$. The carrier concentration is
the calculated density of ions on the $\text{SiO}_{2}$ surface which
is in agreement with the carrier density derived from Hall effect
measurements made at $2\text{K}$ in other Si samples.}
\end{figure}

We can rule out the mobility peak being an electrochemical effect.
The surface oxide passivation layer reduces the trap density on the
Si surface from upwards of $10^{15}\text{ cm}^{-2}$ to values as
low as $10^{11}\text{ cm}^{-2}$ in addition to serving as an insulating
barrier between the ions and conducting channel.\citep{cobbold1970theory}
The carrier density as a function of gate voltage is linear through
the conductance peak consistent with electrostatic rather than electrochemical
behavior. Further support for electrostatic behavior is the fact that
the capacitance calculated from $n$ vs $V_{G}$ from Fig. \ref{fig:4}
agrees with the capacitance expected from a $10\AA$ thick $\text{SiO}_{2}$
capacitor. The gate voltages used are also within the electrochemical
window of DEME-TFSI and so an electrochemical reaction is unlikely.\citep{Yuan(2009)}
The electrochemical window sets an upper limit to the voltage that
when exceeded results in breakdown of the electrolyte.\citep{Ong(2011)}

At high carrier densities, short range interactions dominate the scattering
time as opposed to long range iterations. Das Sarma and Hwang calculated
the scattering time due to surface roughness of the oxide interface,
which is a source of short range disorder, and found that the mobility
would decrease as the sheet carrier density $n$ exceeded $10^{12}\text{ cm}^{-2}$.
The calculation was done to explain the observed mobility peak in
low mobility Si samples. In addition they proposed that for high mobility
Si MOSFETs at sheet carrier densities of around $10^{13}\text{cm}^{-2}$
the mobility would be reduced sufficiently such that $k_{f}l<1$ and
the system would undergo a second MIT and reenter the insulating state.\citep{DasSarma(2014)nc2}

Thus there would be two critical carrier densities, $n_{c1}$ and
$n_{c2}.$ Metallic behavior would be observed when $n_{c1}<n<n_{c2}$.
The most likely value for the carrier density of the second transition
would be $n_{c2}\approx3-5x10^{13}\text{cm}^{-2}$ .\citep{DasSarma(2014)nc2}
Figure \ref{fig:4} shows the mobility calculated from the conductivity
at 2K along with the carrier density as a function of gate voltage.
For this sample, $\sigma\sim1.2\text{ e}^{2}/\text{h}$ from $-1.1\text{V}$
to $-1.3\text{V}$ and in this range the conductivity appears to increase
with decreasing temperature. It is worth noting that the sample plotted
in Fig. \ref{fig:4} was not as strongly metallic as the samples presented
in Fig. \ref{fig:2} or in Fig. 2 of the supplemental material, but
those samples where not pushed into the insulating state by further
increase of the carrier density. A more conclusive result would be
to observe reentrant behavior in a more strongly metallic sample while
increasing carrier densities to drive the sample much further into
the insulating state.

The origin of why we see metallic behavior in a subset of samples
is also an open question. If the peak wafer mobility were to lead
to $n_{c1}>n_{c2}$ it would then appear possible that there would
be no metallic state. This could explain why a metallic regime is
not seen in every sample. Another open question is why there is a
metallic regime in samples whose mobilities are well below those of
Ref. \onlinecite{Ando(1982)} for which a MIT was not observed.
At very high carrier densities screening of Coulomb disorder may play
a role and as a result the observed metallic state may not involve
the same physics as the ones previously seen. 

In summary, ionic liquid gating has been used to induce holes on the
surface of \textit{p}-type \textlangle100\textrangle{} Si wafers.
The contacts were also \textit{p}-type and the samples were cooled
down to freeze out the bulk carriers and thus permit the measurement
of the induced surface conducting channel. The low temperature Hall
mobilities were on the order of $20$$\text{ cm}^{2}\text{/Vs}$ and
not $10^{4}$$\text{ cm}^{2}\text{/Vs}$ as seen in Si MOSFETs that
undergo a MIT. In the low mobility samples, a MIT was found with $n_{c}\sim10^{12}\text{ cm}^{-2},$
a value much higher than that found in all previous studies of the
MIT. It is generally believed that a metallic state would be unlikely
at such a high level of disorder. In addition it was found that above
$n\sim10^{13}\text{ cm}^{-2}$ both metallic and insulating samples
become more insulating with increasing carrier concentration. Finally
the mobility at peak conductance was $3x$ higher than the mobility
peak in rubrene gated with an ionic liquid.\citep{Wei(2014)}
\begin{acknowledgments}
\appendix
This work was supported in part by NSF/DMR-1263316. Devices were fabricated
at the Minnesota Nanofabrication Center which receives partial support
from the NSF through the NNIN program. They were characterized at
the University of Minnesota Characterization Facility, a member of
the NSF-funded Materials Research Facilities Network via the MRSEC
program. The authors would like to thank Boris Shklovskii for useful
discussions.
\end{acknowledgments}


\begin{thebibliography}{22}%
\makeatletter
\providecommand \@ifxundefined [1]{%
 \@ifx{#1\undefined}
}%
\providecommand \@ifnum [1]{%
 \ifnum #1\expandafter \@firstoftwo
 \else \expandafter \@secondoftwo
 \fi
}%
\providecommand \@ifx [1]{%
 \ifx #1\expandafter \@firstoftwo
 \else \expandafter \@secondoftwo
 \fi
}%
\providecommand \natexlab [1]{#1}%
\providecommand \enquote  [1]{``#1''}%
\providecommand \bibnamefont  [1]{#1}%
\providecommand \bibfnamefont [1]{#1}%
\providecommand \citenamefont [1]{#1}%
\providecommand \href@noop [0]{\@secondoftwo}%
\providecommand \href [0]{\begingroup \@sanitize@url \@href}%
\providecommand \@href[1]{\@@startlink{#1}\@@href}%
\providecommand \@@href[1]{\endgroup#1\@@endlink}%
\providecommand \@sanitize@url [0]{\catcode `\\12\catcode `\$12\catcode
  `\&12\catcode `\#12\catcode `\^12\catcode `\_12\catcode `\%12\relax}%
\providecommand \@@startlink[1]{}%
\providecommand \@@endlink[0]{}%
\providecommand \url  [0]{\begingroup\@sanitize@url \@url }%
\providecommand \@url [1]{\endgroup\@href {#1}{\urlprefix }}%
\providecommand \urlprefix  [0]{URL }%
\providecommand \Eprint [0]{\href }%
\providecommand \doibase [0]{http://dx.doi.org/}%
\providecommand \selectlanguage [0]{\@gobble}%
\providecommand \bibinfo  [0]{\@secondoftwo}%
\providecommand \bibfield  [0]{\@secondoftwo}%
\providecommand \translation [1]{[#1]}%
\providecommand \BibitemOpen [0]{}%
\providecommand \bibitemStop [0]{}%
\providecommand \bibitemNoStop [0]{.\EOS\space}%
\providecommand \EOS [0]{\spacefactor3000\relax}%
\providecommand \BibitemShut  [1]{\csname bibitem#1\endcsname}%
\let\auto@bib@innerbib\@empty
\bibitem [{\citenamefont {Abrahams}\ \emph {et~al.}(1979)\citenamefont
  {Abrahams}, \citenamefont {Anderson}, \citenamefont {Licciardello},\ and\
  \citenamefont {Ramakrishnan}}]{Abrahams(1979)}%
  \BibitemOpen
  \bibfield  {author} {\bibinfo {author} {\bibfnamefont {E.}~\bibnamefont
  {Abrahams}}, \bibinfo {author} {\bibfnamefont {P.~W.}\ \bibnamefont
  {Anderson}}, \bibinfo {author} {\bibfnamefont {D.~C.}\ \bibnamefont
  {Licciardello}}, \ and\ \bibinfo {author} {\bibfnamefont {T.~V.}\
  \bibnamefont {Ramakrishnan}},\ }\href {\doibase 10.1103/PhysRevLett.42.673}
  {\bibfield  {journal} {\bibinfo  {journal} {Phys. Rev. Lett.}\ }\textbf
  {\bibinfo {volume} {42}},\ \bibinfo {pages} {673} (\bibinfo {year}
  {1979})}\BibitemShut {NoStop}%
\bibitem [{\citenamefont {Kravchenko}\ \emph {et~al.}(1994)\citenamefont
  {Kravchenko}, \citenamefont {Kravchenko}, \citenamefont {Furneaux},
  \citenamefont {Pudalov},\ and\ \citenamefont {D'Iorio}}]{Kravchenko(1994)}%
  \BibitemOpen
  \bibfield  {author} {\bibinfo {author} {\bibfnamefont {S.~V.}\ \bibnamefont
  {Kravchenko}}, \bibinfo {author} {\bibfnamefont {G.~V.}\ \bibnamefont
  {Kravchenko}}, \bibinfo {author} {\bibfnamefont {J.~E.}\ \bibnamefont
  {Furneaux}}, \bibinfo {author} {\bibfnamefont {V.~M.}\ \bibnamefont
  {Pudalov}}, \ and\ \bibinfo {author} {\bibfnamefont {M.}~\bibnamefont
  {D'Iorio}},\ }\href {\doibase 10.1103/PhysRevB.50.8039} {\bibfield  {journal}
  {\bibinfo  {journal} {Phys. Rev. B}\ }\textbf {\bibinfo {volume} {50}},\
  \bibinfo {pages} {8039} (\bibinfo {year} {1994})}\BibitemShut {NoStop}%
\bibitem [{\citenamefont {Hanein}\ \emph {et~al.}(1998)\citenamefont {Hanein},
  \citenamefont {Meirav}, \citenamefont {Shahar}, \citenamefont {Li},
  \citenamefont {Tsui},\ and\ \citenamefont {Shtrikman}}]{Hanein(1998)}%
  \BibitemOpen
  \bibfield  {author} {\bibinfo {author} {\bibfnamefont {Y.}~\bibnamefont
  {Hanein}}, \bibinfo {author} {\bibfnamefont {U.}~\bibnamefont {Meirav}},
  \bibinfo {author} {\bibfnamefont {D.}~\bibnamefont {Shahar}}, \bibinfo
  {author} {\bibfnamefont {C.~C.}\ \bibnamefont {Li}}, \bibinfo {author}
  {\bibfnamefont {D.~C.}\ \bibnamefont {Tsui}}, \ and\ \bibinfo {author}
  {\bibfnamefont {H.}~\bibnamefont {Shtrikman}},\ }\href {\doibase
  10.1103/PhysRevLett.80.1288} {\bibfield  {journal} {\bibinfo  {journal}
  {Phys. Rev. Lett.}\ }\textbf {\bibinfo {volume} {80}},\ \bibinfo {pages}
  {1288} (\bibinfo {year} {1998})}\BibitemShut {NoStop}%
\bibitem [{\citenamefont {Dultz}\ \emph {et~al.}(1998)\citenamefont {Dultz},
  \citenamefont {Jiang},\ and\ \citenamefont {Schaff}}]{Dultz(1998)}%
  \BibitemOpen
  \bibfield  {author} {\bibinfo {author} {\bibfnamefont {S.~C.}\ \bibnamefont
  {Dultz}}, \bibinfo {author} {\bibfnamefont {H.~W.}\ \bibnamefont {Jiang}}, \
  and\ \bibinfo {author} {\bibfnamefont {W.~J.}\ \bibnamefont {Schaff}},\
  }\href {\doibase 10.1103/PhysRevB.58.R7532} {\bibfield  {journal} {\bibinfo
  {journal} {Phys. Rev. B}\ }\textbf {\bibinfo {volume} {58}},\ \bibinfo
  {pages} {R7532} (\bibinfo {year} {1998})}\BibitemShut {NoStop}%
\bibitem [{\citenamefont {Das~Sarma}\ \emph {et~al.}(2014)\citenamefont
  {Das~Sarma}, \citenamefont {Hwang}, \citenamefont {Kechedzhi},\ and\
  \citenamefont {Tracy}}]{DasSarma(2014)Sept}%
  \BibitemOpen
  \bibfield  {author} {\bibinfo {author} {\bibfnamefont {S.}~\bibnamefont
  {Das~Sarma}}, \bibinfo {author} {\bibfnamefont {E.~H.}\ \bibnamefont
  {Hwang}}, \bibinfo {author} {\bibfnamefont {K.}~\bibnamefont {Kechedzhi}}, \
  and\ \bibinfo {author} {\bibfnamefont {L.~A.}\ \bibnamefont {Tracy}},\ }\href
  {\doibase 10.1103/PhysRevB.90.125410} {\bibfield  {journal} {\bibinfo
  {journal} {Phys. Rev. B}\ }\textbf {\bibinfo {volume} {90}},\ \bibinfo
  {pages} {125410} (\bibinfo {year} {2014})}\BibitemShut {NoStop}%
\bibitem [{\citenamefont {Spivak}\ \emph {et~al.}(2010)\citenamefont {Spivak},
  \citenamefont {Kravchenko}, \citenamefont {Kivelson},\ and\ \citenamefont
  {Gao}}]{Spivak(2010)}%
  \BibitemOpen
  \bibfield  {author} {\bibinfo {author} {\bibfnamefont {B.}~\bibnamefont
  {Spivak}}, \bibinfo {author} {\bibfnamefont {S.~V.}\ \bibnamefont
  {Kravchenko}}, \bibinfo {author} {\bibfnamefont {S.~A.}\ \bibnamefont
  {Kivelson}}, \ and\ \bibinfo {author} {\bibfnamefont {X.~P.~A.}\ \bibnamefont
  {Gao}},\ }\href {\doibase 10.1103/RevModPhys.82.1743} {\bibfield  {journal}
  {\bibinfo  {journal} {Rev. Mod. Phys.}\ }\textbf {\bibinfo {volume} {82}},\
  \bibinfo {pages} {1743} (\bibinfo {year} {2010})}\BibitemShut {NoStop}%
\bibitem [{\citenamefont {Bustarret}\ \emph {et~al.}(2006)\citenamefont
  {Bustarret}, \citenamefont {Marcenat}, \citenamefont {Achatz}, \citenamefont
  {Kacmarcik}, \citenamefont {Levy}, \citenamefont {Huxley}, \citenamefont
  {Ortega}, \citenamefont {Bourgeois}, \citenamefont {Blase}, \citenamefont
  {Debarre},\ and\ \citenamefont {Boulmer}}]{Bustarret2006}%
  \BibitemOpen
  \bibfield  {author} {\bibinfo {author} {\bibfnamefont {E.}~\bibnamefont
  {Bustarret}}, \bibinfo {author} {\bibfnamefont {C.}~\bibnamefont {Marcenat}},
  \bibinfo {author} {\bibfnamefont {P.}~\bibnamefont {Achatz}}, \bibinfo
  {author} {\bibfnamefont {J.}~\bibnamefont {Kacmarcik}}, \bibinfo {author}
  {\bibfnamefont {F.}~\bibnamefont {Levy}}, \bibinfo {author} {\bibfnamefont
  {A.}~\bibnamefont {Huxley}}, \bibinfo {author} {\bibfnamefont
  {L.}~\bibnamefont {Ortega}}, \bibinfo {author} {\bibfnamefont
  {E.}~\bibnamefont {Bourgeois}}, \bibinfo {author} {\bibfnamefont
  {X.}~\bibnamefont {Blase}}, \bibinfo {author} {\bibfnamefont
  {D.}~\bibnamefont {Debarre}}, \ and\ \bibinfo {author} {\bibfnamefont
  {J.}~\bibnamefont {Boulmer}},\ }\href {http://dx.doi.org/10.1038/nature05340}
  {\bibfield  {journal} {\bibinfo  {journal} {Nature}\ }\textbf {\bibinfo
  {volume} {444}},\ \bibinfo {pages} {465} (\bibinfo {year}
  {2006})}\BibitemShut {NoStop}%
\bibitem [{\citenamefont {Zhang}\ \emph {et~al.}(2010)\citenamefont {Zhang},
  \citenamefont {Cheng}, \citenamefont {Li}, \citenamefont {Sun}, \citenamefont
  {Wang}, \citenamefont {Zhu}, \citenamefont {He}, \citenamefont {Wang},
  \citenamefont {Ma}, \citenamefont {Chen}, \citenamefont {Wang}, \citenamefont
  {Liu}, \citenamefont {Lin}, \citenamefont {Jia},\ and\ \citenamefont
  {Xue}}]{Zhang2010}%
  \BibitemOpen
  \bibfield  {author} {\bibinfo {author} {\bibfnamefont {T.}~\bibnamefont
  {Zhang}}, \bibinfo {author} {\bibfnamefont {P.}~\bibnamefont {Cheng}},
  \bibinfo {author} {\bibfnamefont {W.-J.}\ \bibnamefont {Li}}, \bibinfo
  {author} {\bibfnamefont {Y.-J.}\ \bibnamefont {Sun}}, \bibinfo {author}
  {\bibfnamefont {G.}~\bibnamefont {Wang}}, \bibinfo {author} {\bibfnamefont
  {X.-G.}\ \bibnamefont {Zhu}}, \bibinfo {author} {\bibfnamefont
  {K.}~\bibnamefont {He}}, \bibinfo {author} {\bibfnamefont {L.}~\bibnamefont
  {Wang}}, \bibinfo {author} {\bibfnamefont {X.}~\bibnamefont {Ma}}, \bibinfo
  {author} {\bibfnamefont {X.}~\bibnamefont {Chen}}, \bibinfo {author}
  {\bibfnamefont {Y.}~\bibnamefont {Wang}}, \bibinfo {author} {\bibfnamefont
  {Y.}~\bibnamefont {Liu}}, \bibinfo {author} {\bibfnamefont {H.-Q.}\
  \bibnamefont {Lin}}, \bibinfo {author} {\bibfnamefont {J.-F.}\ \bibnamefont
  {Jia}}, \ and\ \bibinfo {author} {\bibfnamefont {Q.-K.}\ \bibnamefont
  {Xue}},\ }\href {http://dx.doi.org/10.1038/nphys1499} {\bibfield  {journal}
  {\bibinfo  {journal} {Nat Phys}\ }\textbf {\bibinfo {volume} {6}},\ \bibinfo
  {pages} {104} (\bibinfo {year} {2010})}\BibitemShut {NoStop}%
\bibitem [{\citenamefont {Das~Sarma}\ and\ \citenamefont
  {Hwang}(2014{\natexlab{a}})}]{DasSarma(2014)2D}%
  \BibitemOpen
  \bibfield  {author} {\bibinfo {author} {\bibfnamefont {S.}~\bibnamefont
  {Das~Sarma}}\ and\ \bibinfo {author} {\bibfnamefont {E.~H.}\ \bibnamefont
  {Hwang}},\ }\href {\doibase 10.1103/PhysRevB.89.235423} {\bibfield  {journal}
  {\bibinfo  {journal} {Phys. Rev. B}\ }\textbf {\bibinfo {volume} {89}},\
  \bibinfo {pages} {235423} (\bibinfo {year} {2014}{\natexlab{a}})}\BibitemShut
  {NoStop}%
\bibitem [{\citenamefont {Misra}\ \emph {et~al.}(2007)\citenamefont {Misra},
  \citenamefont {McCarthy},\ and\ \citenamefont {Hebard}}]{APLMisra(2007)}%
  \BibitemOpen
  \bibfield  {author} {\bibinfo {author} {\bibfnamefont {R.}~\bibnamefont
  {Misra}}, \bibinfo {author} {\bibfnamefont {M.}~\bibnamefont {McCarthy}}, \
  and\ \bibinfo {author} {\bibfnamefont {A.~F.}\ \bibnamefont {Hebard}},\
  }\href {\doibase http://dx.doi.org/10.1063/1.2437663} {\bibfield  {journal}
  {\bibinfo  {journal} {Applied Physics Letters}\ }\textbf {\bibinfo {volume}
  {90}},\ \bibinfo {eid} {052905} (\bibinfo {year} {2007})}\BibitemShut
  {NoStop}%
\bibitem [{\citenamefont {Shimotani}\ \emph {et~al.}(2007)\citenamefont
  {Shimotani}, \citenamefont {Asanuma}, \citenamefont {Tsukazaki},
  \citenamefont {Ohtomo}, \citenamefont {Kawasaki},\ and\ \citenamefont
  {Iwasa}}]{Shimotani(2007)}%
  \BibitemOpen
  \bibfield  {author} {\bibinfo {author} {\bibfnamefont {H.}~\bibnamefont
  {Shimotani}}, \bibinfo {author} {\bibfnamefont {H.}~\bibnamefont {Asanuma}},
  \bibinfo {author} {\bibfnamefont {A.}~\bibnamefont {Tsukazaki}}, \bibinfo
  {author} {\bibfnamefont {A.}~\bibnamefont {Ohtomo}}, \bibinfo {author}
  {\bibfnamefont {M.}~\bibnamefont {Kawasaki}}, \ and\ \bibinfo {author}
  {\bibfnamefont {Y.}~\bibnamefont {Iwasa}},\ }\href {\doibase
  http://dx.doi.org/10.1063/1.2772781} {\bibfield  {journal} {\bibinfo
  {journal} {Applied Physics Letters}\ }\textbf {\bibinfo {volume} {91}},\
  \bibinfo {eid} {082106} (\bibinfo {year} {2007})}\BibitemShut {NoStop}%
\bibitem [{\citenamefont {Jeong}\ \emph {et~al.}(2013)\citenamefont {Jeong},
  \citenamefont {Aetukuri}, \citenamefont {Graf}, \citenamefont {Schladt},
  \citenamefont {Samant},\ and\ \citenamefont {Parkin}}]{Jeong22032013}%
  \BibitemOpen
  \bibfield  {author} {\bibinfo {author} {\bibfnamefont {J.}~\bibnamefont
  {Jeong}}, \bibinfo {author} {\bibfnamefont {N.}~\bibnamefont {Aetukuri}},
  \bibinfo {author} {\bibfnamefont {T.}~\bibnamefont {Graf}}, \bibinfo {author}
  {\bibfnamefont {T.~D.}\ \bibnamefont {Schladt}}, \bibinfo {author}
  {\bibfnamefont {M.~G.}\ \bibnamefont {Samant}}, \ and\ \bibinfo {author}
  {\bibfnamefont {S.~S.~P.}\ \bibnamefont {Parkin}},\ }\href {\doibase
  10.1126/science.1230512} {\bibfield  {journal} {\bibinfo  {journal}
  {Science}\ }\textbf {\bibinfo {volume} {339}},\ \bibinfo {pages} {1402}
  (\bibinfo {year} {2013})}\BibitemShut {NoStop}%
\bibitem [{\citenamefont {Borghesi}\ \emph {et~al.}(1995)\citenamefont
  {Borghesi}, \citenamefont {Pivac}, \citenamefont {Sassella},\ and\
  \citenamefont {Stella}}]{Borghesi(1995)}%
  \BibitemOpen
  \bibfield  {author} {\bibinfo {author} {\bibfnamefont {A.}~\bibnamefont
  {Borghesi}}, \bibinfo {author} {\bibfnamefont {B.}~\bibnamefont {Pivac}},
  \bibinfo {author} {\bibfnamefont {A.}~\bibnamefont {Sassella}}, \ and\
  \bibinfo {author} {\bibfnamefont {A.}~\bibnamefont {Stella}},\ }\href
  {\doibase http://dx.doi.org/10.1063/1.359479} {\bibfield  {journal} {\bibinfo
   {journal} {Journal of Applied Physics}\ }\textbf {\bibinfo {volume} {77}},\
  \bibinfo {pages} {4169} (\bibinfo {year} {1995})}\BibitemShut {NoStop}%
\bibitem [{Sup()}]{SupDoc}%
  \BibitemOpen
  \href {To be inserted by the publisher} {}\bibinfo {note} {See Supplemental
  Material at [URL will be inserted by publisher]}\BibitemShut {NoStop}%
\bibitem [{\citenamefont {Ando}\ \emph {et~al.}(1982)\citenamefont {Ando},
  \citenamefont {Fowler},\ and\ \citenamefont {Stern}}]{Ando(1982)}%
  \BibitemOpen
  \bibfield  {author} {\bibinfo {author} {\bibfnamefont {T.}~\bibnamefont
  {Ando}}, \bibinfo {author} {\bibfnamefont {A.~B.}\ \bibnamefont {Fowler}}, \
  and\ \bibinfo {author} {\bibfnamefont {F.}~\bibnamefont {Stern}},\ }\href
  {\doibase 10.1103/RevModPhys.54.437} {\bibfield  {journal} {\bibinfo
  {journal} {Rev. Mod. Phys.}\ }\textbf {\bibinfo {volume} {54}},\ \bibinfo
  {pages} {437} (\bibinfo {year} {1982})}\BibitemShut {NoStop}%
\bibitem [{\citenamefont {Petach}\ \emph {et~al.}(2014)\citenamefont {Petach},
  \citenamefont {Lee}, \citenamefont {Davis}, \citenamefont {Mehta},\ and\
  \citenamefont {Goldhaber-Gordon}}]{Petach(2014)}%
  \BibitemOpen
  \bibfield  {author} {\bibinfo {author} {\bibfnamefont {T.~A.}\ \bibnamefont
  {Petach}}, \bibinfo {author} {\bibfnamefont {M.}~\bibnamefont {Lee}},
  \bibinfo {author} {\bibfnamefont {R.~C.}\ \bibnamefont {Davis}}, \bibinfo
  {author} {\bibfnamefont {A.}~\bibnamefont {Mehta}}, \ and\ \bibinfo {author}
  {\bibfnamefont {D.}~\bibnamefont {Goldhaber-Gordon}},\ }\href {\doibase
  10.1103/PhysRevB.90.081108} {\bibfield  {journal} {\bibinfo  {journal} {Phys.
  Rev. B}\ }\textbf {\bibinfo {volume} {90}},\ \bibinfo {pages} {081108}
  (\bibinfo {year} {2014})}\BibitemShut {NoStop}%
\bibitem [{\citenamefont {Zavaritskaya}\ and\ \citenamefont
  {Zavaritskaya}(1987)}]{Zataritskaya(1987)}%
  \BibitemOpen
  \bibfield  {author} {\bibinfo {author} {\bibfnamefont {T.~N.}\ \bibnamefont
  {Zavaritskaya}}\ and\ \bibinfo {author} {\bibfnamefont {E.~I.}\ \bibnamefont
  {Zavaritskaya}},\ }\href@noop {} {\bibfield  {journal} {\bibinfo  {journal}
  {JETP Letters}\ }\textbf {\bibinfo {volume} {45}},\ \bibinfo {pages} {609}
  (\bibinfo {year} {1987})}\BibitemShut {NoStop}%
\bibitem [{\citenamefont {Xie}\ \emph {et~al.}(2014)\citenamefont {Xie},
  \citenamefont {Wang}, \citenamefont {Zhang}, \citenamefont {Leighton},\ and\
  \citenamefont {Frisbie}}]{Wei(2014)}%
  \BibitemOpen
  \bibfield  {author} {\bibinfo {author} {\bibfnamefont {W.}~\bibnamefont
  {Xie}}, \bibinfo {author} {\bibfnamefont {S.}~\bibnamefont {Wang}}, \bibinfo
  {author} {\bibfnamefont {X.}~\bibnamefont {Zhang}}, \bibinfo {author}
  {\bibfnamefont {C.}~\bibnamefont {Leighton}}, \ and\ \bibinfo {author}
  {\bibfnamefont {C.~D.}\ \bibnamefont {Frisbie}},\ }\href {\doibase
  10.1103/PhysRevLett.113.246602} {\bibfield  {journal} {\bibinfo  {journal}
  {Phys. Rev. Lett.}\ }\textbf {\bibinfo {volume} {113}},\ \bibinfo {pages}
  {246602} (\bibinfo {year} {2014})}\BibitemShut {NoStop}%
\bibitem [{\citenamefont {Cobbold}(1970)}]{cobbold1970theory}%
  \BibitemOpen
  \bibfield  {author} {\bibinfo {author} {\bibfnamefont {R.}~\bibnamefont
  {Cobbold}},\ }\href {http://books.google.com/books?id=ISVTAAAAMAAJ} {\emph
  {\bibinfo {title} {Theory and applications of field-effect transistors}}}\
  (\bibinfo  {publisher} {Wiley-Interscience},\ \bibinfo {year}
  {1970})\BibitemShut {NoStop}%
\bibitem [{\citenamefont {Yuan}\ \emph {et~al.}(2009)\citenamefont {Yuan},
  \citenamefont {Shimotani}, \citenamefont {Tsukazaki}, \citenamefont {Ohtomo},
  \citenamefont {Kawasaki},\ and\ \citenamefont {Iwasa}}]{Yuan(2009)}%
  \BibitemOpen
  \bibfield  {author} {\bibinfo {author} {\bibfnamefont {H.}~\bibnamefont
  {Yuan}}, \bibinfo {author} {\bibfnamefont {H.}~\bibnamefont {Shimotani}},
  \bibinfo {author} {\bibfnamefont {A.}~\bibnamefont {Tsukazaki}}, \bibinfo
  {author} {\bibfnamefont {A.}~\bibnamefont {Ohtomo}}, \bibinfo {author}
  {\bibfnamefont {M.}~\bibnamefont {Kawasaki}}, \ and\ \bibinfo {author}
  {\bibfnamefont {Y.}~\bibnamefont {Iwasa}},\ }\href {\doibase
  10.1002/adfm.200801633} {\bibfield  {journal} {\bibinfo  {journal} {Advanced
  Functional Materials}\ }\textbf {\bibinfo {volume} {19}},\ \bibinfo {pages}
  {1046} (\bibinfo {year} {2009})}\BibitemShut {NoStop}%
\bibitem [{\citenamefont {Ong}\ \emph {et~al.}(2011)\citenamefont {Ong},
  \citenamefont {Andreussi}, \citenamefont {Wu}, \citenamefont {Marzari},\ and\
  \citenamefont {Ceder}}]{Ong(2011)}%
  \BibitemOpen
  \bibfield  {author} {\bibinfo {author} {\bibfnamefont {S.~P.}\ \bibnamefont
  {Ong}}, \bibinfo {author} {\bibfnamefont {O.}~\bibnamefont {Andreussi}},
  \bibinfo {author} {\bibfnamefont {Y.}~\bibnamefont {Wu}}, \bibinfo {author}
  {\bibfnamefont {N.}~\bibnamefont {Marzari}}, \ and\ \bibinfo {author}
  {\bibfnamefont {G.}~\bibnamefont {Ceder}},\ }\href {\doibase
  10.1021/cm200679y} {\bibfield  {journal} {\bibinfo  {journal} {Chemistry of
  Materials}\ }\textbf {\bibinfo {volume} {23}},\ \bibinfo {pages} {2979}
  (\bibinfo {year} {2011})},\ \Eprint
  {http://arxiv.org/abs/http://dx.doi.org/10.1021/cm200679y}
  {http://dx.doi.org/10.1021/cm200679y} \BibitemShut {NoStop}%
\bibitem [{\citenamefont {Das~Sarma}\ and\ \citenamefont
  {Hwang}(2014{\natexlab{b}})}]{DasSarma(2014)nc2}%
  \BibitemOpen
  \bibfield  {author} {\bibinfo {author} {\bibfnamefont {S.}~\bibnamefont
  {Das~Sarma}}\ and\ \bibinfo {author} {\bibfnamefont {E.~H.}\ \bibnamefont
  {Hwang}},\ }\href {\doibase 10.1103/PhysRevB.89.121413} {\bibfield  {journal}
  {\bibinfo  {journal} {Phys. Rev. B}\ }\textbf {\bibinfo {volume} {89}},\
  \bibinfo {pages} {121413} (\bibinfo {year} {2014}{\natexlab{b}})}\BibitemShut
  {NoStop}%
\end{thebibliography}

%

\end{document}